\documentclass[conference, 10pt]{IEEEtran}
\usepackage[dvips]{graphicx}
\usepackage{amssymb,amsmath}
\usepackage{subfig}
\usepackage{colortbl}
\usepackage{url}
\usepackage{cite}
\usepackage{algorithm2e}
\usepackage{multirow}

\makeatletter
\def\ps@headings{%
\def\@oddhead{\mbox{}\scriptsize\rightmark \hfil \thepage}%
\def\@evenhead{\scriptsize\thepage \hfil \leftmark\mbox{}}%
\def\@oddfoot{}%
\def\@evenfoot{}}
\makeatother
\IEEEoverridecommandlockouts

\begin{document}

\title{CTCP: Coded TCP using Multiple Paths}
\author{
\authorblockN{MinJi Kim, Ali ParandehGheibi, Leonardo Urbina, Muriel M\'edard}\vspace*{-.7cm}
\thanks{The authors are with the RLE at the Massachusetts Institute of Technology, MA USA (e-mail: \{minjikim, parandeh, lurbina, medard\}@mit.edu).}
}

\maketitle

\begin{abstract}
We introduce CTCP, a novel multi-path transport protocol using network coding. CTCP is designed to incorporate TCP's good features, such as congestion control and reliability, while improving on TCP's performance in lossy and/or dynamic networks. CTCP builds upon the ideas of TCP/NC introduced by Sundararajan et al. and uses network coding to provide robustness against losses. We introduce the use of multiple paths to provide robustness against mobility and network failures. We provide an implementation of CTCP (in userspace) to demonstrate its performance.
\end{abstract}
\IEEEpeerreviewmaketitle

\section{Introduction}\label{sec:introduction}

The Transmission Control Protocol (TCP) is one of the core protocols of today's Internet Protocol Suite. TCP was designed for reliable transmission over wired networks, in which losses are generally indication of congestion. This is not the case in wireless networks, where losses are often due to fading, interference, and other physical phenomena. Consequently, TCP's performance in wireless networks is poor when compared to the wired counterparts as shown e.g. in \cite{caceres, TCP_Kurose}. There has been extensive research to combat these harmful effects of erasures and failures \cite{sack, Tian05tcpin, stcp}; however, TCP even with modifications does not achieve significant improvement. References \cite{hari, Tian05tcpin} give an overview and a comparison of various TCP versions over wireless links.

In recent years, there has been proposals to use multiple paths available on devices to enhance TCP's performance (in terms of robustness and throughput). New congestion control and scheduling algorithms have been proposed to support multiple paths with TCP (MPTCP) \cite{mptcp,mptcp2}. Enabling multiple paths allow for a device to receive continuous data connection while moving between and across multiple networks and access technology. However, MPTCP like TCP suffers in performance when there are losses.

In this paper, we introduce CTCP, a novel multi-path transport protocol using \emph{network coding} \cite{ahlswede,xor,more,codeOr,costa,uusee}, which has been introduced as a potential paradigm to operate communication networks in particular wireless networks. Network coding allows and encourages mixing of data at intermediate nodes, which has been shown to increase throughput and robustness against failures and erasures \cite{algebraic}. In order to combine the benefits of TCP and network coding, \cite{tcpnc} proposes a new protocol called TCP/NC. TCP/NC modifies TCP's acknowledgment (ACK) scheme so that it acknowledges \emph{degrees of freedom} instead of individual packets. This is done so by using the concept of ``seen'' packets -- in which the number of degrees of freedom received is translated to the number of consecutive packets received. Reference \cite{analysis} provides mathematical model and analysis with simulation results that show that TCP/NC achieves significantly higher throughput in lossy networks.

The goal of our design is to keep traditional TCP's best features, including congestion control and reliability, and augment them with network coding and the use of multiple paths for resilience against losses and failures in the network. We uphold the end-to-end philosophy of TCP, and only require the sender and the receiver to change. Therefore, our main contributions are as follows.
\begin{enumerate}
\item We present an algorithm for multi-path communication. We show that CTCP is able to achieve a combined throughput of the available paths. The key challenge in allowing for multi-path connection is in designing an algorithm which intelligently assigns traffic over multiple paths. CTCP uses round-trip time (RTT) estimates, packet loss rates, and the throughput of the different paths to assign traffic over multiple paths.
\item We implement the design in C for Linux operating systems. We implement the protocol in userspace (over UDP) for ease of implementation and modifications. The implementation demonstrates that CTCP is indeed able to achieve high throughput despite losses in the network and achieve a combined throughput of multiple paths.
\end{enumerate}

Our protocol, CTCP, builds upon TCP/NC introduced by Sundararajan et al. \cite{tcpnc}. Note that TCP/NC does not enable multiple paths. We use many of the ideas from TCP/NC to design the algorithms for each path in CTCP. However, we enhance the algorithms from TCP/NC to make each path in CTCP more efficient and robust. Therefore, even in a single path case, CTCP departs from TCP/NC in the following ways.
\begin{enumerate}
\item CTCP, unlike TCP/NC which introduces network coding indirectly through a shim layer between TCP and IP layers, is a transport protocol that uses network coding directly. At a first glance, this may not be a significant change; however, by designing a new transport protocol, we are able to leverage network coding more efficiently and introduce congestion control mechanism better suited to a protocol using network coding.
\item CTCP is adaptive. TCP/NC assumes a known average end-to-end packet loss rate $p$ and determines a redundancy factor $R \sim \frac{1}{1-p}$ for the communication \cite{tcpnc}. Note that this redundancy factor plays a key role in enabling TCP/NC to use network coding to overcome losses. However, in a real network, $p$ is rarely known a priori and fluctuates over time and space. Our protocol estimates $p$ and dynamically adjusts the redundancy rate $R$ to adjust to the losses on the fly.
\item CTCP uses systematic block coding to manage delay and complexity. TCP/NC uses a sliding window approach for coding operations, which can have significant decoding delay at the receiver as it may have undesirable worst-case behavior.
Furthermore, the use of a systematic code significantly reduces the decoding overhead over a random solutions with dense matrix. When there are no losses ($p = 0$ leading to $R \sim 1$), CTCP in effect reduces to traditional TCP without coding.
\item We provide a congestion control mechanism works well with network coding. Traditional TCP uses a sliding transmission window with sequence numbers to identify bytes. With coding, any coded packet within a block can replace another packet; therefore, we modify the congestion control mechanism to use tokens -- i.e. a token allows CTCP sender to transmit a packet. Our congestion control mechanism generates or destroys tokens to adjust CTCP sender's transmission rate. The congestion control mechanism uses round-trip time estimates (similarly to TCP-Vegas \cite{vegas,analsys_low}) as well as packet loss rates to adjust the number of tokens.
\end{enumerate}

This paper discusses the algorithmic and implementation details of our proposed protocol. The rest of the paper is organized as follows. In Section \ref{sec:overview}, we provide an overview of CTCP. In Sections \ref{sec:sender} and \ref{sec:receiver}, we describe CTCP sender and receiver, respectively. Before presenting the experimental results in Section \ref{sec:experiments}, we present some of the implementation details in Section \ref{sec:implementation}. Finally, we conclude in Section \ref{sec:conclusions}.

\section{Overview of CTCP}\label{sec:overview}

\begin{figure}[tbp]
\begin{center}\vspace*{.4cm}
\includegraphics[width=.47\textwidth]{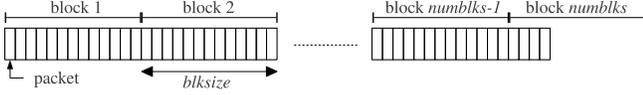}
\end{center}\vspace*{-.3cm}
\caption{CTCP sender divides the file or stream into blocks.}
\label{fig:blocks}\vspace*{-.4cm}
\end{figure}

 CTCP sender segments the stream or the file into blocks as shown in Figure \ref{fig:blocks}. A block is chosen to be of a fixed size, equivalent to $blksize$ number of packets where each packet is assumed to be of fixed length. If the remainder of a file or a stream is not large enough to form a complete packet, the packet is padded with zeros to ensure that all packets are of the same length. A block need not be completely full, i.e. a block may have fewer than $blksize$ packets; however, block $i$ should be full before block $i+1$ is initialized.

 CTCP sender keep $numblks$ of blocks in memory, and the value of $numblks$ should be conveyed to the receiver. The value of $numblks$ may be negotiated at initialization between the sender and the receiver, as $numblks$ directly affect the memory usage on both ends. We denote the smallest block in memory to be $currblk$. Note that this does not mean that CTCP sender may send $numblks\times blksize$ amount of data at any given point in time. The sender is allowed to transmit packets only if the congestion control mechanism allows it to; however, whenever it is allowed to transmit, the sender may choose to transmit a packet from any one of the blocks in memory, i.e. blocks $currblk$, $currblk+1$, ..., $currblk+numblks-1$. In Section \ref{sec:blockschedule}, we shall discuss the sender's algorithm for selecting a block to transmit from. The payload of the transmitted packet may be coded or uncoded; the details of the coding operations can be found in Section \ref{sec:coding}.

The sender includes in a packet
\begin{itemize}
\item the block number,
\item coding coefficients,
\item the sequence number $seqno$, and
\item the (coded or uncoded) payload.
\end{itemize}
Note that sequence number for CTCP differs from that of TCP -- for TCP, a sequence number indicates a specific data byte; for CTCP, a sequence number indicates that a packet is the $seqno$-th packet transmitted by the sender, thus, is not tied to a byte in the file. As we shall discuss in Section \ref{sec:estimation}, $seqno$ is used to estimate losses and round-trip times. It is important to realize that losses and round-trip times have to be estimated for each path separately, as different paths may display different characteristics. As a result, when multiple paths are used for a CTCP connection, the sender uses a distinct series of sequence numbers for the different paths. Therefore, we introduce the notation $seqno_i$ to denote the sequence number of the packets transmitted on path $i$.

As we shall discuss in Section \ref{sec:receiver}, the receiver sends acknowledgments (ACKs) for the packets it received. In the ACK, the receiver indicates
\begin{itemize}
\item the smallest undecoded block $ack\_currblk$,
\item the number of degrees of freedom (dofs) $ack\_currdof$ it has received from the current block $ack\_currblk$, and
\item the $ack\_seqno$ of the packet it is acknowledging.
\end{itemize}

Using the information carried in an ACK, CTCP sender adjusts its behavior. We first describe the sender-side algorithm in Section \ref{sec:sender}, as most of the intelligence of the protocol is on the sender's side. Then, we present CTCP receiver algorithm in Section \ref{sec:receiver}. CTCP receiver's main role is to decode and deliver data in order to the application.

\section{CTCP Sender}\label{sec:sender}

We present the source side algorithm for CTCP. CTCP sender maintains several internal parameters as shown in Table \ref{tab:defintions}, which it uses to generate coded packets and schedule blocks to be transmitted.

\begin{table}[tbp]
\caption{Notations and definitions of the sender parameters}\label{tab:defintions}
\begin{tabular}{|l|p{6.0cm}|}
\hline
Notation & Definition\\
\hline
\hline
$p_i$ & Short term average packet loss rate for path $i$\\
$p\_long_i$ & Long term average packet loss rate for path $i$\\
$p\_stdlong_i$ & Long term standard deviation of the packet loss rate for path $i$\\
$RTO_i$ & Retransmission timeout period (equal to $\gamma\cdot RTT_i$ where $\gamma\geq 1$ is a constant)\\
$RTT_i$ & Short term average round-trip time for path $i$\\
$seqno\_nxt_i$ & The sequence number of the next packet to be transmitted on path $i$\\
$seqno\_una_i$ & The sequence number of the latest unacknowledged packet on path $i$\\
$ss\_threshold_i$ & Slow-start threshold for path $i$, i.e. if $tokens_i > ss\_threshold_i$, the sender leaves the slow-start mode on path $i$\\
$time\_lastack_i$ & Timestamp of when the sender received the latest ACK on path $i$ (initialized to the time when the sender receives a SYN packet from the receiver)\\
$tokens_i$ & Number of tokens for path $i$, which is conceptually similar to congestion window for traditional TCP\\
$blksize$ & Size of the blocks (in number of packets)\\
$currblk$ & Current block number at the sender, which is the smallest unacknowledged block number\\
$currdof$ & Number of dofs the receiver has acknowledged for the current block\\
$numblks$ & Number of active blocks, i.e. the sender may schedule and transmit packets from blocks $currblk$, $currblk+1$, ..., $currblk+numblks-1$\\
$B(seqno)$ & Block number from which packet with $seqno$ was generated from\\
$T(seqno)$ & Timestamp of when packet with $seqno$ was sent\\
\hline
\end{tabular}
\end{table}

\subsection{Network Parameter Estimation}\label{sec:estimation}

CTCP sender estimates the network parameters, such as $RTT_i$, $p_i$, and $p\_long_i$, using the ACKs as shown in Algorithm \ref{alg:update}. The sender adjusts its actions, including coding operations and congestion control depending on the values of the network parameters. If the sender does not receive an ACK from the receiver for an extended period of time (i.e. time-out occurs), the network parameters are reset to predefined default values. The predefined default values may need to be chosen with some care such that they estimate roughly what the network may look like.

CTCP sender maintains moving averages of the parameters. For $RTT_i$, we use an exponential smoothing technique. However, for $p_i$ and $p\_long_i$, we use a slightly modified version of the exponential smoothing technique. For $p_i$ and $p\_long_i$, we can consider the data series we are averaging to be a 0-1 sequence, where 0 indicates that the packet has been sent successfully and 1 otherwise. Now, assume that there were $losses$ number of packets lost. If $losses = 0$, then the update equation for $p_i$ in Algorithm \ref{alg:update} becomes
\begin{equation}
p_i \leftarrow p_i(1-\mu)+0.
\end{equation}
If $losses = 1$, the same update equation becomes
\begin{equation}
p_i\leftarrow p_i(1-\mu)^{2}+\mu= (1-\mu)[p_i (1-\mu) + 0 ] + \mu,
\end{equation}
which is identical to executing an exponential smoothing over two data points (one lost and one acknowledged). We can repeat this idea for $losses > 1$ to obtain the update rule for $p_i$ in Algorithm \ref{alg:update}. Therefore, the fact that a single ACK may represent multiple losses ($losses\geq 1$) leads to a slightly more complicated update rule for $p_i$ than that for $RTT_i$ as shown in Algorithm \ref{alg:update}. To the best of our knowledge, such an update rule has not been used previously. The same logic applies to $p\_long_i$. Note that $p_i$ has a shorter memory than $p\_long_i$; therefore, $\nu < \mu$.

\begin{algorithm}[tbp]
Receive an ACK on path $i$\;
$time\_lastack_i \leftarrow$ current time\;
$rtt = time\_lastack_i - T(ack\_seqno)$\;
$RTT_i \leftarrow RTT_i\cdot (1-\alpha) + rtt\cdot \alpha$\;
\If{$ack\_currblk > currblk$}
{
Free blocks $currblk$, ..., $ack\_currblk-1$\;
$currdof \leftarrow ack\_currdof$\;
$currblk \leftarrow ack\_currblk$\;
}
\If{$ack\_seqno \geq seqno\_una_i$}
{
$losses = ack\_seqno - seqno\_una_i$\;
$p_i \leftarrow p_i(1-\mu)^{losses+1}+(1-(1-\mu)^{losses})$\;
$p\_long_i \leftarrow p\_long_i(1-\nu)^{losses+1}+(1-(1-\nu)^{losses})$\;
$p\_stdlong_i \leftarrow p\_stdlong_i(1-\nu)+\nu\cdot |p_i - p\_long_i|$\;
}
$seqno\_una_i \leftarrow ack\_seqno+1$\;
$currdof \leftarrow \max\{ack\_currdof, currdof\}$\;
\vspace*{.2cm}
\caption{CTCP sender algorithm for updating the network parameters.}\label{alg:update}
\end{algorithm}

\subsection{Reliability}\label{sec:reliability}

CTCP achieves reliability by ensuring that each block is received and decoded. In Algorithm \ref{alg:update}, CTCP sender increments $currblk$ only if it has received an ACK indicating that the receiver is able to decode $currblk$ -- i.e. $ack\_currblk > currblk$. This mechanism is equivalent to traditional TCP's window sliding scheme in which the TCP sender only slides its window when it receives an ACK indicating the some bytes have been received. In the case of CTCP, the reliability is implemented over blocks instead of bytes.

\subsection{Congestion Control Mechanism}\label{sec:congestion}

The congestion control for CTCP is similar to TCP-Vegas \cite{vegas}; however, is adapted to suit the need of CTCP as shown in Algorithm \ref{alg:congestion}. We use \emph{tokens} instead of congestion window to control CTCP sender's transmission rate. A token allows CTCP sender to transmit a packet (coded or uncoded), and when the sender transmits a packet, the token is used. Token regeneration is managed as follows:
\begin{itemize}
\item when an ACK is received, then a token is regenerated;
\item if losses are detected (i.e. $losses > 0$), then $losses$ number of tokens are regenerated;
\item if timeout occurs, then a small fixed number of tokens are regenerated; at this point, any ACKs for packets sent prior to the timeout has no effect.
\end{itemize}

A timeout occurs when the sender does not receive any ACKs for more than $RTO_i$ time period. CTCP uses the slow-start mechanism to ramp up the number of tokens; however, after a certain threshold, CTCP enters the congestion avoidance mode. During congestion avoidance mode, CTCP uses $RTT_i$, similar to TCP-Vegas \cite{vegas}, to control its transmission rate. In addition, CTCP uses loss rate estimates $p_i$. When there is a large increase in loss rate (i.e. $p_i > p\_long_i + p\_stdlong_i$), CTCP uses this as a sign of congestion and reduces its rate. This latter method is specific to CTCP.

\begin{algorithm}[tbp]
\If{\emph{current time} $> time\_lastack_i + RTO_i$}
{
$RTO_i \leftarrow 2\cdot RTO_i$\;
$ss\_threshold_i \leftarrow \frac{tokens_i}{2}$\;
$tokens_i \leftarrow$ initial token number\;
$seqno\_una_i \leftarrow seqno\_nxt_i$\;
Set path $i$ to slow-start mode\;
}
\If{\emph{Receive an ACK on path} $i$}{
$rtt \leftarrow time\_lastack_i - T(ack\_seqno)$\;
$RTO_i \leftarrow \gamma \cdot RTT_i$\;
\eIf{\emph{path $i$ is in slow-start mode}}
{
$tokens_i \leftarrow tokens_i + 1$\;
\If{$tokens_i > ss\_threshold_i$}
{
Set path $i$ to congestion avoidance mode\;
}
}
{
$\delta \leftarrow 1- \frac{RTT_i}{rtt}$\;
\uIf{$\delta > \beta$}
{
$tokens_i \leftarrow tokens_i - \frac{1}{tokens_i}$\;
}\ElseIf{$\delta < \alpha$}{
$tokens_i \leftarrow tokens_i + \frac{1}{tokens_i}$\;
}
}

\If{$p_i > p\_long_i + p\_stdlong_i$}
{
$tokens_i \leftarrow tokens_i - \frac{p_i - p\_long_i}{2}$\;
}
}
\vspace*{.2cm}
\caption{CTCP sender algorithm for congestion control mechanism.}\label{alg:congestion}
\end{algorithm}

\subsection{Coding Operations}\label{sec:coding}

The coding operations are performed over blocks (Figure \ref{fig:blocks}). Unlike TCP/NC \cite{tcpnc}, we do not use a sliding window for coding operations. The main reason behind this design decision is for delay and complexity. The sliding window approach allows for better throughput performance; however, when using this approach, the receiver may not be able to decode even the first packet of the file until the entire file is received. As a result, the decoding complexity may be high, as the decoding operation may have to perform over the entire file (instead of segments of the file). This may not be a significant concern for small file transfers; however, for some applications such as multimedia streaming and large file transfers, this may be a significant concern. Therefore, in our design, we have opted to use block codes, where we can bound the delay and the complexity by changing the block size $blksize$.

Setting $blksize = 1$ leads to operations similar to that of traditional TCP variants and cannot effectively take advantage of network coding. On the other hand, setting $blksize$ too large leads to the problems faced by TCP/NC. Therefore, $blksize$ has to be chosen with care.  In our experience, it is desirable to set $blksize$ to be similar to the bandwidth$\times$delay of the network. This is because if $blksize$ is too small, the sender may potentially be sending many blocks of data in an open-loop -- i.e. without any feedback from the receiver. This may be a problem, especially when sender is either unaware of or unable to adapt to the changes in the network.

To ensure that coding is only performed when necessary, we use systematic block codes -- i.e. uncoded packets are transmitted before coded packets are sent. In generating coded packets, there are many options. The sender may only code a subset of the packets in a block. In our design, we use a simple approach -- a coded packet is generated by randomly coding all packets in the block together. This approach is most effective in terms erasure correction. With high probability, a coded packet will correct for any single erasure in the block.

\subsection{Transmission and Block Scheduling}\label{sec:blockschedule}

When a token is available, CTCP sender decides which block to transmit a packet from. The block scheduling algorithm (Algorithm \ref{alg:block}) plays a key role in CTCP's operations.

The algorithm first computes the number of packets in transit from the sender to the receiver on path $i$. Given $p_i$, the sender can compute the expected number of packets the receiver will receive for any given block. In determining the expected number of dofs the receiver will receive for any given block, we exclude the packets that have been transmitted more than $1.5\cdot RTT_i$ time ago, as they are likely to be lost or significantly delayed. The constant factor of 1.5 may be adjusted depending on the delay constraints of the application of interest; however, the constant factor should be $\geq 1$.

The goal of the sender is to ensure that, in expectation, the receiver will receive enough packets to decode the block. The sender prioritizes block $i$ before $i+1$; therefore, $currblk$ is of the highest priority. Note that the algorithm treats $currblk$ slightly differently from the rest of the blocks. In our design, CTCP receiver informs the sender of how many dofs it has received ($currdof$) for block $currblk$. Therefore, the sender is able to use the additional information to determine more precisely whether another packet should be sent from block $currblk$ or not. It is not difficult to piggy-back more information on the ACKs. For example, we could include how many dofs the receiver has received for blocks $currblk$ as well as $currblk+1$, $currblk+2$, ..., $currblk+numblks-1$. However, for simplicity, CTCP receiver only informs the sender the number of dofs received for block $currblk$.

In Algorithm \ref{alg:block}, we assume that all blocks are of length $blksize$. We note that CTCP can cope with blocks of varying length; however, for simplicity of presentation, we have chosen to present the algorithms with a fixed block length.

\begin{algorithm}[tbp]
Initialize an array $onfly[]$ to $0$\;
\For{$seqno$ \emph{in} $[seqno\_una_i, seqno\_nxt_i -1 ]$}
{
\If{\emph{current time} $< T(seqno) + 1.5 RTT_i$}
{
$onfly[B(seqno)] \leftarrow onfly[B(seqno)] + 1$\;
}
}
\For{$blkno$ \emph{in} $[currblk, currblk+numblks-1]$}{
\uIf{$blkno = currblk$ \emph{and} $(1-p_i)onfly[currblk] <  blksize - currdof$}
{
Transmit a packet with sequence number $seqno\_nxt_i$ from block $blkno$\;
$seqno\_nxt_i \leftarrow seqno\_nxt_i + 1$\;
break\;
}\ElseIf{$(1-p_i)onfly[blkno] < blksize$}{
Transmit a packet with sequence number $seqno\_nxt_i$ from block $blkno$\;
$seqno\_nxt_i \leftarrow seqno\_nxt_i + 1$\;
break\;
}
}
\vspace*{.2cm}
\caption{CTCP sender algorithm for block scheduling for a single path $i$.}\label{alg:block}
\end{algorithm}

\subsection{Multi-path Scheduling}\label{sec:multipath}

When multiple paths are available between the sender and the receiver, CTCP protocol may take advantage of them. CTCP performs coding operations independent of the multiple paths. For each path, CTCP sender estimates independently a set of network parameters (as discussed in Section \ref{sec:estimation}). Furthermore, the congestion control on each path is independent -- i.e. the number of tokens $tokens_i$ for path $i$ does not affect the number of tokens $token_j$ for path $j$, where $i \ne j$. This design decision was made to prevent low bandwidth paths to slow down the entire connection.

However, in order to take advantage efficiently of the multiple paths, we need to modify Algorithm \ref{alg:block} because of the following scenario. Imagine two paths (1 and 2) with different characteristics: e.g. $RTT_1 \ll RTT_2$. The discrepancies between the two paths may cause path 2 delaying the entire connection. Assume that CTCP sender sends a packet from $currblk$ on path 2. When a token becomes available for path 1, the sender now needs to decide whether it should send from $currblk$ or not. If the sender chooses to send from $currblk$ on path 1, the packet sent over path 2 may become redundant; thus resulting in lower throughput but the receiver may be able to decode $currblk$ sooner. Depending on how the sender decides to handle these cases, we may observe significant improvement or degradation in performance. A bad decision at the sender may lead result in a slow connection despite having a high bandwidth path available.

We propose the block scheduling algorithm in Algorithm~\ref{alg:multipath} to address this problem. The first for-loop computes the number of packets in flight per block. For example, $thru$ computes the combined average throughput of the connection (over all paths), and $sent$ computes the number of packets in flight that will be received by the receiver in expectation. The variables $cof_{blkno}$ computes the number of packets from block $blkno$ in flight that will be received by the receiver.

Given these estimates, CTCP sender then computes which block to transmit a packet from (the second for-loop). CTCP sender gives preference to lower block numbers, with the highest priority given to $currblk$. For $currblk$, CTCP sender takes into account the delay in its path (i.e. $thru\cdot RTT_i$); thus, if the given path $i$ has a large delay, it may choose to not send a packet from $currblk$ as another path $j \ne i$ where $RTT_j < RTT_i$ may be able to handle $currblk$ more efficiently. If the sender evaluates that path $i$ is suitable for sending a packet from $currblk$, the sender checks whether in expectation the receiver will receive enough packets from $currblk$, and if not, transmits a packet from $currblk$. Otherwise, the sender moves on to other blocks. For $blkno > currblk$, CTCP sender does not take into account $RTT_i$ but only checks the number of packets in flight. This was a design decision we made to simplify the algorithm at the sender and the structure of the ACKs. However, it is not difficult to modify and extend CTCP to use more sophisticated decision policies for blocks $blkno > currblk$.

Note that CTCP sender may choose not to transmit a packet although a token is available on a given path. In this case, CTCP sender reserves the token for future use. It is not difficult to extend CTCP sender so that whenever a token becomes available it transmits a packet (preferably from $currblk$). Such design may lead to a more aggressive CTCP sender; however, may come at the cost of transmitting more redundant packets. We chose to be conservative and allow CTCP sender to defer the use of a token.

\begin{algorithm}[tbp]
Initialize $thru$ and $sent$ to 0\;
Initialize an array $onfly_k[]$ for each path $k$ to 0\;
Initialize $cof_{blkno} = 0$ for each active block $blkno$\;
\For{$k$ \emph{in paths}}{
\For{$seqno$ \emph{in} $[seqno\_una_k, seqno\_nxt_k -1 ]$}
{
\If{\emph{current time} $> T(seqno) + 1.5 RTT_k$}
{
$onfly_k[B(seqno)] \leftarrow onfly_k[B(seqno)] + 1$\;
}
}
$thru \leftarrow thru + (1-p_k)\frac{seqno\_nxt_k - seqno\_una_k}{RTT_k}$\;
$sent \leftarrow sent + (1-p_k)(seqno\_nxt_k - seqno\_una_k)$\;
\For{$blkno$ in $[currblk, currblk+numblks-1]$}
{
$cof_{blkno} \leftarrow cof_{blkno} + (1-p_k)onfly_k[blkno]$\;
}
}
\For{$blkno$ in $[currblk, currblk+numblks-1]$}
{
\uIf{$blkno = currblk$ \emph{and} $thru\cdot RTT_i - sent + cof_{currblk} < currdof$}
{
Send a packet with sequence number $seqno\_nxt_i$ from block $currblk$\;
$seqno\_nxt_i \leftarrow seqno\_nxt_i + 1$\;
return\;
}\ElseIf{$cof_{blkno} < blksize$ }
{
Send a packet with sequence number $seqno\_nxt_i$ from block $blkno$\;
$seqno\_nxt_i \leftarrow seqno\_nxt_i + 1$\;
return\;
}
}
\vspace*{.2cm}
\caption{CTCP sender algorithm for block scheduling over multiple paths for path $i$.
}\label{alg:multipath}
\end{algorithm}

\section{CTCP Receiver}\label{sec:receiver}

We now present the receiver side algorithm for CTCP. The receiver is responsible for decoding the received data. Another important role of the receiver is to construct acknowledgments (ACKs) for the sender. Whenever the receiver receives a packet, it needs to check whether the current block is decodable ($ack\_currblk$) and how many dofs it has received for the current block ($ack\_currdof$).

\subsection{Decoding Operations}\label{sec:decoding}

\begin{algorithm}[tbp]
$index \leftarrow$ index of the first non-zero element in $c$\;
\uIf{$C_{blkno}[index, :]$ \emph{is empty}}
{
$pivot \leftarrow$ value of $c$ at index $index$\;
Insert $c/pivot$ into $C_{blkno}[index, :]$\;
Insert $p/pivot$ into $P_{blkno}[index, :]$\;
return TRUE\;
}
\Else{
\If{$index < blksize$}
{
$pivot \leftarrow$ value of $c$ at index $index$\;
$c \leftarrow c - pivot\cdot C_{blkno}[index, :]$\;
$p \leftarrow p - pivot\cdot P_{blkno}[index, :]$\;
$pivot \leftarrow$ value of $c$ at index $index+1$\;
$c \leftarrow c/pivot$\;
$p \leftarrow p/pivot$\;
\If{$c\ne 0$}
{
Recursively call itself with updated $c$ and $p$\;
}
}
return FALSE\;
}
\vspace*{.2cm}
\caption{CTCP receiver algorithm for updating $C_{blkno}$ and $P_{blkno}$ when a packet from block $blkno$ is received. We denote $c$ to be the coding coefficients and $p$ the (coded) payload of the received packet.
}\label{alg:insert}
\end{algorithm}

CTCP receiver also organizes the received packets into blocks. For each block $blkno$, the receiver initializes a $blksize\times blksize$ matrix $C_{blkno}$ for the coding coefficients and a corresponding payload structure $P_{blkno}$. Whenever a packet from $blkno$ is received, the coding coefficients and the coded payload are inserted to $C_{blkno}$ and $P_{blkno}$ respectively as shown in Algorithm \ref{alg:insert}. Algorithm \ref{alg:insert} returns FALSE if the packet is linearly dependent to the previously received packets; otherwise it returns TRUE. Note that Algorithm \ref{alg:insert} ensures that $C_{blkno}$ is an upper-triangular matrix with diagonal entries equal to one. Since CTCP sender uses a systematic code, CTCP receiver may often be able to insert $p$ and $c$ directly -- i.e. row $index$ of $C_{blkno}$ is empty.

When a packet is received and Algorithm \ref{alg:insert} returns TRUE, then CTCP receiver sends an ACK with increment $ack\_currdof\leftarrow ack\_currdof+1$. If $ack\_currdof = blksize$, then the receiver can acknowledge that enough dofs have been received for $ack\_currblk$ and update $ack\_currblk \leftarrow ack\_currblk+1$ (and reset $ack\_currdof$ to reflect the dofs needed for the new $ack\_currblk$). If Algorithm \ref{alg:insert} returns FALSE, then the receiver transmits an ACK (corresponding to the packet receiver); however, does not update $ack\_currdof$ nor $ack\_currblk$.

Once enough dofs are received for a given block, the receiver now can decode the block. This results in performing a Gauss-Jordan elimination on a upper-triangular matrix $C_{blkno}$ and its corresponding $P_{blkno}$.

\section{Implementation Details}\label{sec:implementation}

We implement CTCP protocol in userspace over UDP. This enables us to test CTCP protocol without having to modify the kernel's protocol stack. Our implementation performs all the functionalities described in Sections \ref{sec:overview}, \ref{sec:sender}, and \ref{sec:receiver}. In this section, we discuss some of these details, which may be of interest to the readers.

\subsection{Proxy}\label{sec:proxy}

\begin{figure}[tbp]
\begin{center}
\subfloat[Setup using TCP (without any modification)]{\label{fig:proxy_noproxy}\includegraphics[width=.47\textwidth]{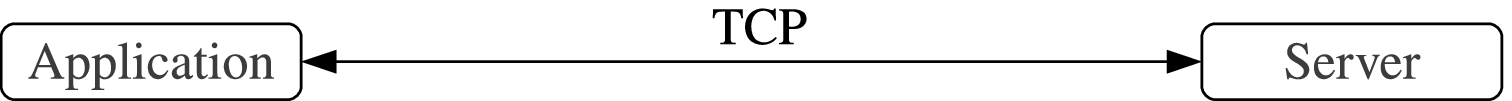}}\\
\subfloat[Setup using CTCP (uses proxies)]{\label{fig:proxy_proxy}\includegraphics[width=.47\textwidth]{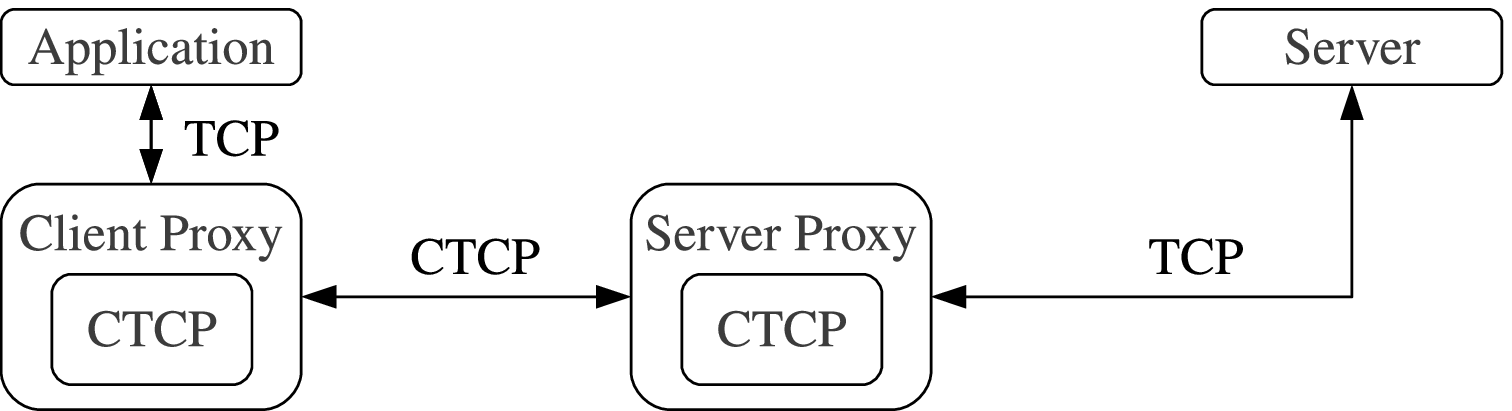}}
\end{center}\vspace*{-.2cm}
\caption{Traditionally, an application and a server communicate directly via a TCP connection. In order to use and test CTCP with existing applications, such as a web browser, we introduce the use of proxy as shown in Figure \ref{fig:proxy}.}
\label{fig:proxy}
\end{figure}

In our design of CTCP, we have introduced the notion of blocks and the algorithms associated with the blocks. The benefit of using blocks is particularly exemplified in delay-sensitive applications, such as multimedia streaming. However, many streaming servers (such as YouTube) are not readily available for us to modify. Therefore, we introduce the use of proxies to enable the use of CTCP as shown in Figure \ref{fig:proxy}.

We can consider applications such as a web browser, which already have customizable network proxy settings. The application connects to the Client Proxy, which hosts CTCP client, using proxy protocols such as SOCKS. The Client Proxy then forwards the request from the application to the Server Proxy. Note that the communication protocol between the Client Proxy and the Server Proxy may be any custom protocol; of course, in this paper, we use CTCP. The Server Proxy sends the request to the server and fetches the data on behalf of the application. The data then travels from the server to the application via the Server Proxy and the Client Proxy.

\subsection{Enabling Multi-path}\label{sec:multihome}

\begin{table*}
\caption{The commands for adding {\tt iptables} to enable multiple interfaces. The parameters are shown in {\tt <>}. {\tt <k>} is the table number; therefore, for each interface, should assign a different number. {\tt <xx.xx.xx.xx>} is the IP address of the interface, {\tt <itf>} is the interface name, {\tt <gg.gg.gg.gg>} is the gateway IP address, {\tt <yy.yy.yy.yy>} is the result of bit-wise AND operation of {\tt <xx.xx.xx.xx>} and the subnet-mask value, and {\tt <zz>} is the subnet-mask number.}\label{tab:iptables}
\begin{verbatim}
ip route add table <k> <yy.yy.yy.yy>/<zz> dev <itf> proto static src <xx.xx.xx.xx>
ip route add table <k> default via <gg.gg.gg.gg> dev <itf> proto static
iptables -t mangle -A OUTPUT -s <xx.xx.xx.xx> -j MARK --set-mark <k>
ip rule add fwmark <k> table <k>
\end{verbatim}
\end{table*}
\begin{table*}
\caption{The commands for adding artificial loss rate of {\tt <p>}.}\label{tab:losses}
\begin{verbatim}
iptables -A INPUT -m statistic --mode random --probability <p> -j DROP
\end{verbatim}
\end{table*}

In order to enable multi-path, CTCP client machine needs to send and receive packets over multiple interfaces. In general, many operating systems and network protocols are designed with the assumption that there is a single primary interface and possibly many other secondary back-up interfaces. In our case, we would like to have multiple interfaces work simultaneously. This is a typical multi-homing problem, which may have many different solutions.

In our setup, we use {\tt iptables}, a tool for IP filtering and NAT. For each interface, we create a new set of filtering/routing rules called a {\tt table}. A {\tt table} is created using dhcp lease information, which can be found in {\tt /var/lib/dchp}. From a dhcp lease, we use information {\tt interface} (the name of the interface such as {\tt eth0}, {\tt wlan0}, etc.), {\tt fixed-address} (the IP address of the interface), {\tt option subnet-mask}, and {\tt option routers} (the interface's gateway IP address).

Given this information, the commands shown in Table \ref{tab:iptables} adds and configures the IP filtering (on the client machine) to enable multiple interfaces. Once the connection terminates and there is no need to have multiple interfaces active, the tables should be deleted and flushed. We do not show the commands for removing the tables for brevity.

In addition to the {\tt iptables}, it may be necessary to set the kernel network parameters appropriately. For example, many systems will have {\tt rp\_filter} (reversed path filter) on, which is recommended for single homed hosts. However, to enable multiple interfaces, {\tt rp\_filter = 0} may be necessary.

\subsection{Injecting Losses}\label{sec:losses}
When operating over WiFi, even in an uncrowded area, losses may occur. However, the losses may occur in bursts and fluctuate over time, making it difficult to run consistent experiments and tests. As a result, we inject losses artificially to test both TCP and CTCP under lossy conditions. To inject losses, we again use the {\tt iptables} module as in Table \ref{tab:losses}. In order to remove the artificial packet drops, we use the same command as shown in Table \ref{tab:losses} with {\tt -A} replaced with {\tt -D}.

\section{Experimental Results}\label{sec:experiments}

\begin{figure*}[tbp]
\begin{center}
\subfloat[{\tt <p>}=0]{\includegraphics[width=0.3\textwidth]{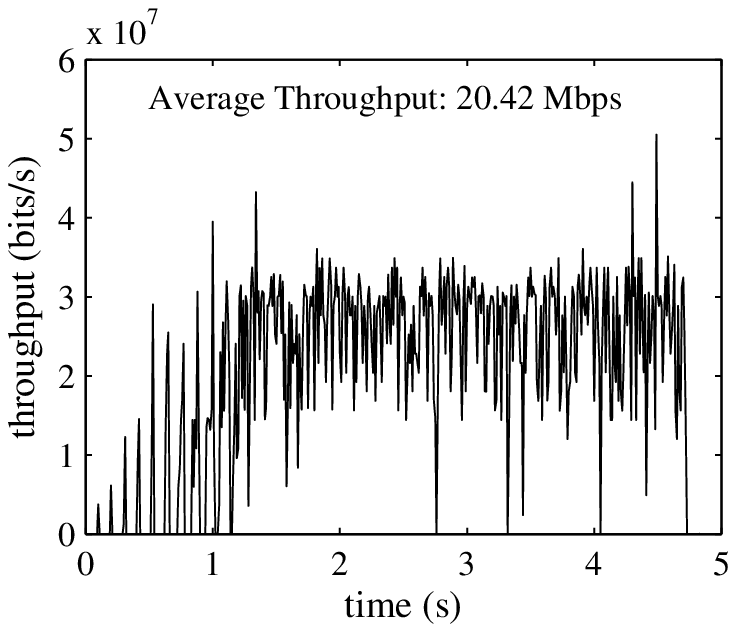}\label{fig:tcp_0}}
\subfloat[{\tt <p>}=0.02]{\includegraphics[width=0.3\textwidth]{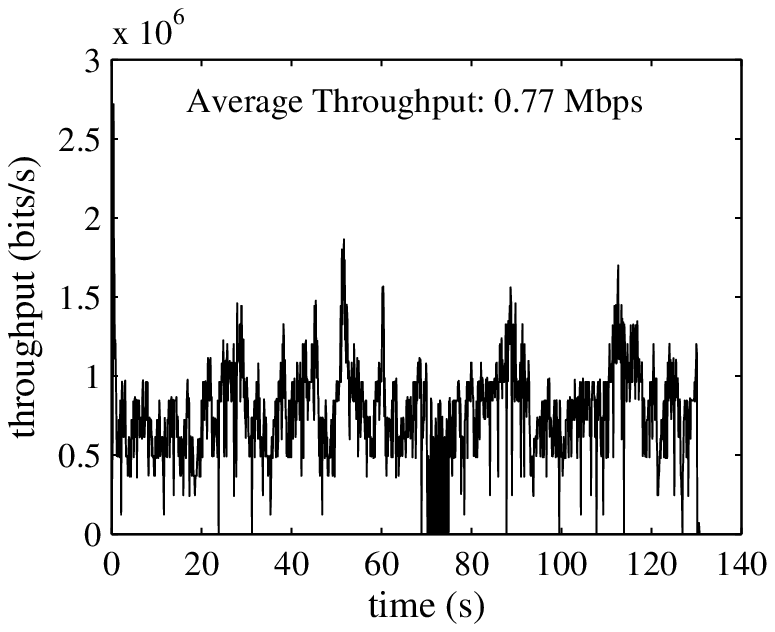}\label{fig:tcp_2}}
\subfloat[{\tt <p>}=0.04]{\includegraphics[width=0.3\textwidth]{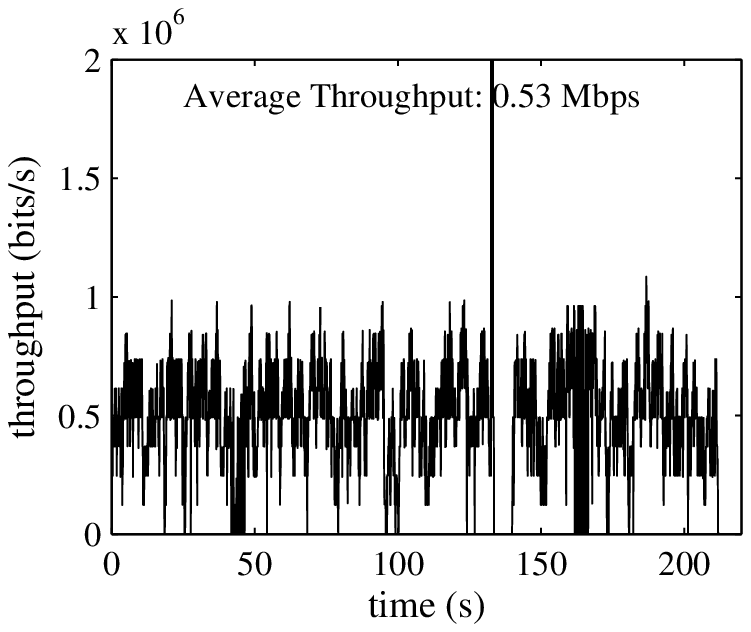}\label{fig:tcp_4}}
\end{center}\vspace*{-.2cm}\caption{Throughput of TCP with varying {\tt <p>}. Note the increase in the duration of the transaction, from 4 seconds when {\tt <p>}=0 to almost 200 seconds when {\tt <p>}=0.04.}\label{fig:tcp}\vspace*{-.5cm}
\end{figure*}

\begin{figure*}[tbp]
\begin{center}
\subfloat[{\tt <p>}=0]{\includegraphics[width=0.3\textwidth]{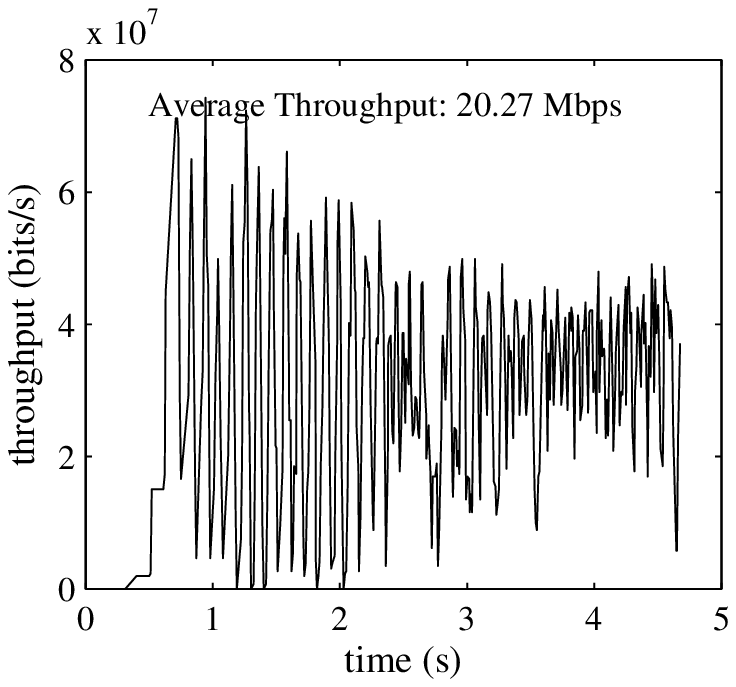}\label{fig:ctcp_0}}
\subfloat[{\tt <p>}=0.02]{\includegraphics[width=0.3\textwidth]{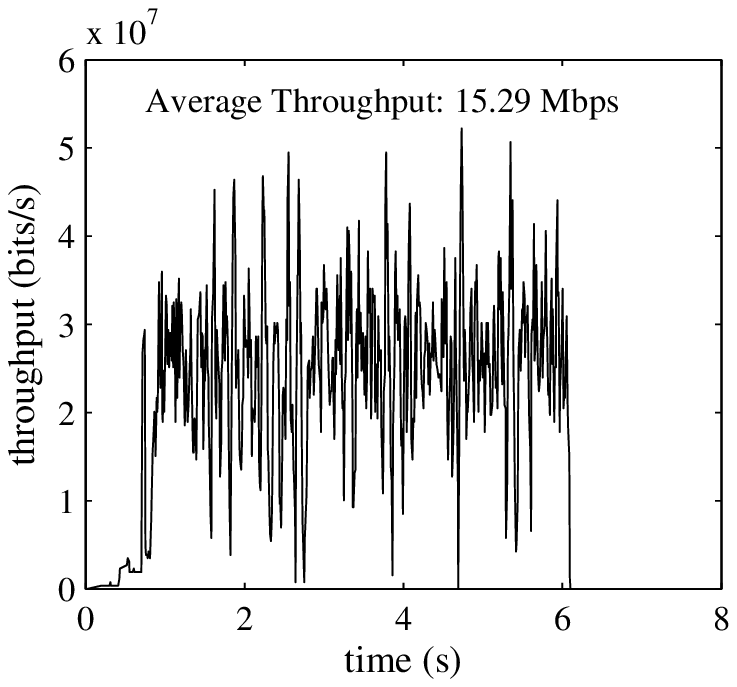}\label{fig:ctcp_2}}
\subfloat[{\tt <p>}=0.04]{\includegraphics[width=0.3\textwidth]{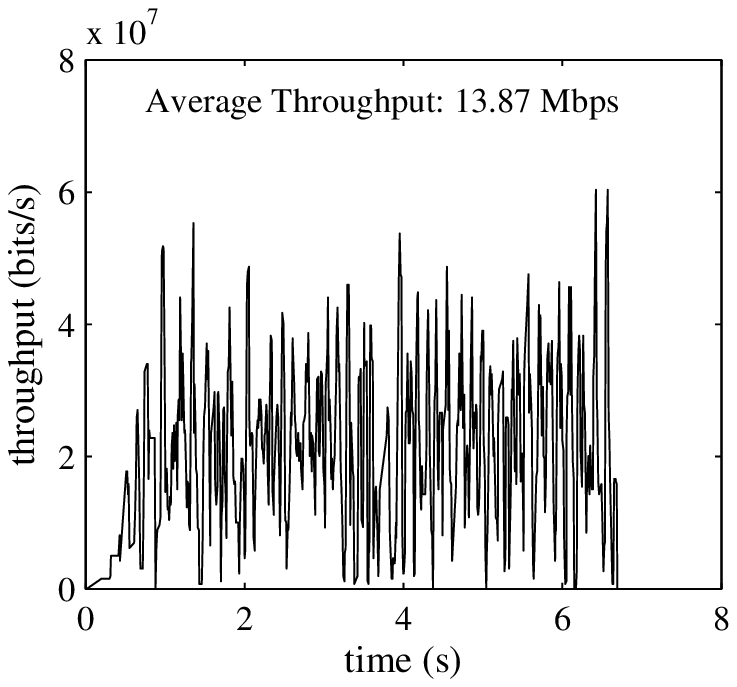}\label{fig:ctcp_4}}
\end{center}\vspace*{-.2cm}\caption{Throughput of CTCP with varying {\tt <p>}. The duration of the transaction stays relatively small, only varying from 4 seconds to 7 seconds.}\label{fig:ctcp}\vspace*{-.4cm}
\end{figure*}

\begin{figure*}[tbp]
\begin{center}
\subfloat[{\tt wlan0}]{\includegraphics[width=0.3\textwidth]{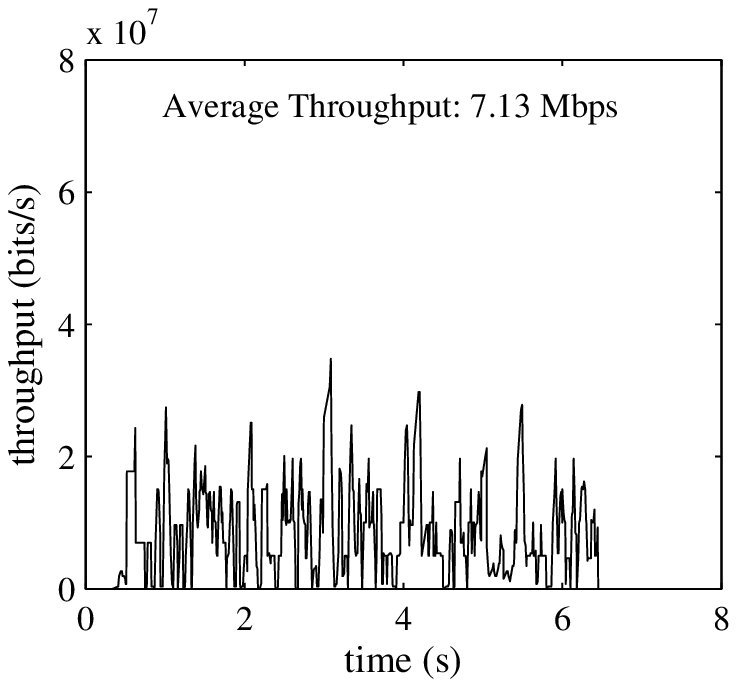}\label{fig:multi_00}}
\subfloat[{\tt wlan1}]{\includegraphics[width=0.3\textwidth]{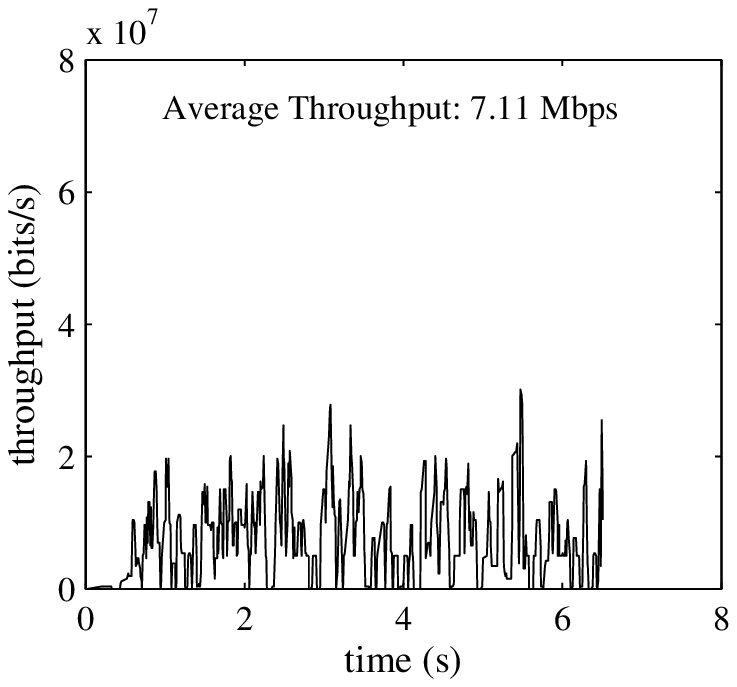}\label{fig:multi_01}}
\subfloat[Total throughput]{\includegraphics[width=0.3\textwidth]{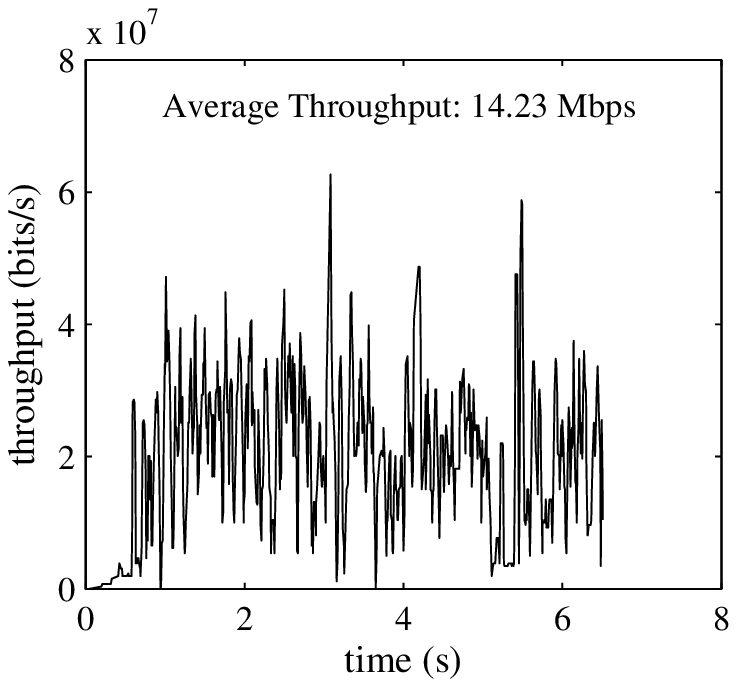}\label{fig:multi_0c}}
\end{center}\vspace*{-.2cm}\caption{Throughput of CTCP with {\tt <p>=0} using two paths ({\tt wlan0} and {\tt wlan1}).}\label{fig:multi_0}\vspace*{-.5cm}
\end{figure*}

\begin{figure*}[tbp]
\begin{center}
\subfloat[{\tt wlan0}]{\includegraphics[width=0.3\textwidth]{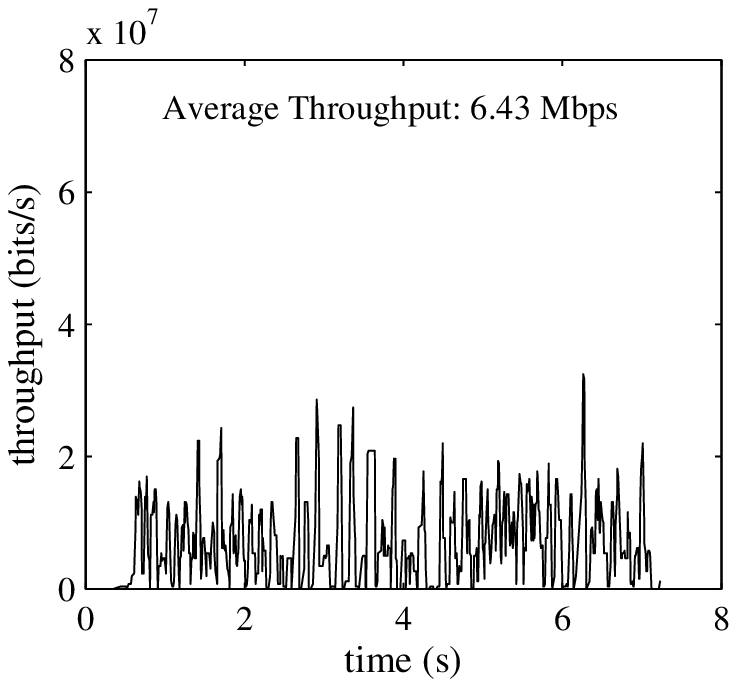}\label{fig:multi_20}}
\subfloat[{\tt wlan1}]{\includegraphics[width=0.3\textwidth]{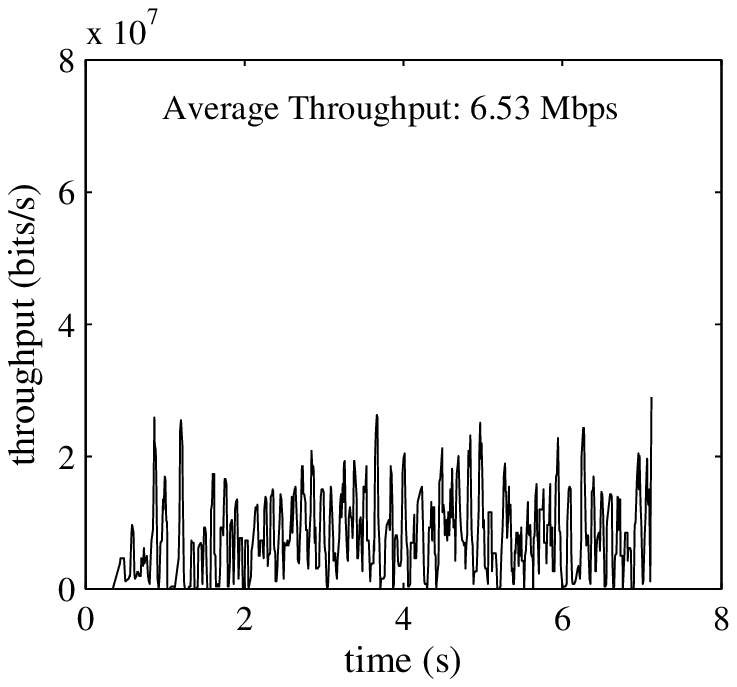}\label{fig:multi_21}}
\subfloat[Total throughput]{\includegraphics[width=0.3\textwidth]{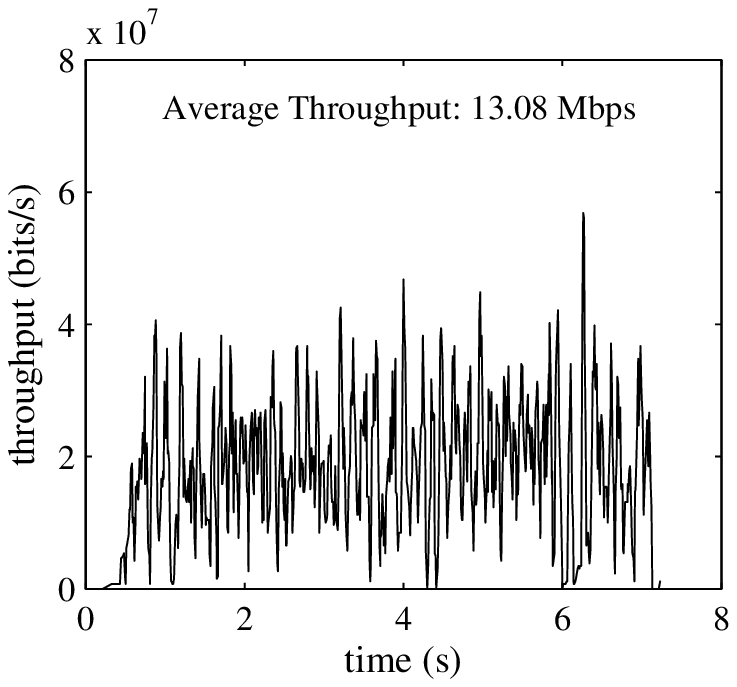}\label{fig:multi_2c}}
\end{center}\vspace*{-.2cm}\caption{Throughput of CTCP with {\tt <p>=0.02} using two paths ({\tt wlan0} and {\tt wlan1}).}\label{fig:multi_2}\vspace*{-.4cm}
\end{figure*}

We present some results demonstrating that indeed CTCP achieves significantly better performance than TCP in faulty or lossy wireless networks. Furthermore, we present results that indicate that CTCP is able to achieve the combined throughput of multiple interfaces. In this paper, we only consider the case with two WiFi interfaces; however, CTCP (at least conceptually) may support many more interfaces simultaneously.

For TCP connections, we use an FTP server ({\tt vsftpd}). We use the default configuration of TCP on our Linux server; therefore, we use CUBIC TCP with SACK enabled. For CTCP connections, we run CTCP server as described previously. We place both the FTP server and CTCP server on an Amazon EC2 micro instance. For the client machine, we use a desktop with Core i3 550 3.2 Ghz processor running Ubuntu 11.10. We use one (for single path CTCP) or two WiFi dongles (for multi-path CTCP) on this desktop to test and run the protocols over the wireless link(s). The RTT between the server and the client is approximately 100 ms.

The results in this paper are driven from a file-transfer application. For all experiments, we transfer a file of 11,492,499 bytes ($\approx$ 11 MB). However, using the proxy setup described in Section \ref{sec:proxy}, we have also tested CTCP for multimedia streaming applications such as YouTube. Similar performance gains can be observed for multimedia streaming (better throughput, which translates to fewer interrupts in the playback).

\subsection{Performance}\label{sec:performance}

As shown in Table \ref{tab:results}, CTCP maintains a significantly higher throughput despite the losses injected. When there are no losses (i.e. {\tt <p>}=0), we observe that CTCP and TCP perform similarly with an average of approximately 4.5 seconds to complete the file transfer. However, the performance of the two protocols diverge quickly as we inject more losses.

\begin{table}[tbp]
\caption{The average performance of TCP and CTCP with varying loss rate {\tt <p>} (using Table \ref{tab:losses}). The results are averaged over 5 runs of FTP/TCP sessions and CTCP sessions.}\label{tab:results}
\centering
\begin{tabular}{|c|c c|c c|}
\hline
\multirow{2}{*}{{\tt <p>}} & \multicolumn{2}{c|}{TCP} & \multicolumn{2}{c|}{CTCP}\\
& duration (s) & Mbps & duration (s) & Mbps\\
\hline
0 & 4.8 & 19.04 &  4.5 & 20.40\\
0.01 & 31.0 & 3.07 & 4.9 & 18.99\\
0.02 & 87.8 & 1.07 & 5.7 & 16.52\\
0.03 & 114.8 & 0.82 & 6.2 & 15.20\\
0.04 & 168.0 & 0.53 & 7.2 & 13.46\\
0.05 & 194.0 & 0.47 & 7.0 & 13.53\\
\hline
\end{tabular}
\end{table}

We also present a more detailed results with another set of experiments. The plots in Figures \ref{fig:tcp} and \ref{fig:ctcp} show an instance of the behavior of TCP and CTCP, respectively. When {\tt <p>}=0, both CTCP and TCP achieve an average throughput of approximately 20 Mbps. In this specific instance, as shown in Figure \ref{fig:ctcp_0}, CTCP experienced a large variance in the beginning of the connection, which could potentially be caused by multiple external network conditions; however, soon converges to 20 Mbps. Note that as {\tt <p>} increases, CTCP maintains its high throughput; completing the entire file download within approximately 7 seconds even when {\tt <p>}=0.04. However, for TCP, even with {\tt <p>}=0.01, the duration of the transaction increases to almost 100 seconds. When {\tt <p>}=0.04, it takes TCP almost 4 minutes to download an 11 MB file.

\subsection{Multiple Interfaces}\label{sec:twowifi}

\begin{table}[tbp]
\caption{The average performance of multi-path CTCP with varying loss rate {\tt <p>}. The results are averaged over 5 runs. We show the throughput achieved by single-path and two-path CTCP for comparison. All the measurements are in Mbps.}\label{tab:multi_results}
\centering
\begin{tabular}{|c|c|c c c|}
\hline
\multirow{2}{*}{{\tt <p>}} & \multirow{2}{*}{Single-path CTCP} & \multicolumn{3}{c|}{Multi-path CTCP}\\
& & {\tt wlan0} & {\tt wlan1} & combined\\
\hline
0 & 7.75 & 6.71 & 6.70 & 13.60\\
0.01 & 7.80 & 6.15 & 6.12 & 12.39\\
0.02 & 7.75 & 6.23 & 6.27 & 12.57\\
0.03 & 6.09 & 5.32 & 5.25 & 10.45\\
0.04 & 6.62 & 4.95 & 4.96 & 9.71\\
0.05 & 6.26 & 4.57 & 4.68 & 10.90\\
\hline
\end{tabular}
\end{table}

In order to demonstrate how CTCP may aggregate multiple interfaces, we attach two WiFi dongles to the client machine. We limit maximum window size on each of the path to approximately 100 packets worth of bytes. This leads to a maximum throughput per path to be at most 7-8 Mbps per path. We limit the maximum throughput per path as the total available bandwidth from our Amazon EC2 server is limited to approximately 20 Mbps (the throughput achieved by TCP and CTCP in Section \ref{sec:performance}).

In Table \ref{tab:multi_results}, we present the average performance of CTCP over single path and multiple paths. In the case with single path, CTCP's performance maintains a throughput of approximately 6-8 Mbps regardless of the loss rate. For multi-path CTCP, we observe a total throughput equal to the sum of the throughput of the two paths. In Figures \ref{fig:multi_0} and \ref{fig:multi_2}, we observe that the traffic is generally well divided between the two paths ({\tt wlan0} and {\tt wlan1}). However, we observe that as loss rates increase, the throughput achieved by each path decreases noticeably more. Note that the decrease is still not as dramatic as TCP (even when it is not throttled).

We believe that the degradation of CTCP's performance over multi-path is due to the multi-path scheduling algorithm described in Section \ref{sec:multipath}. The algorithm attempts to schedule traffic across multiple paths without too much redundancy; however, as loss rates increase, CTCP sender's ability to predict and appropriately assign traffic are hindered. We believe that there are still room for significant improvement in the multi-path scheduling algorithm. Our algorithm is relatively simple, only using the first moment of the network parameters. Despite its limitations, we believe that our algorithm is one of the first to use coding for multi-path communications. The biggest benefit of coding in multi-path communication is that it alleviates the need to schedule carefully each individual byte/packet to each path (e.g. MPTCP \cite{mptcp,mptcp2}); CTCP allows us to assign portions of the blocks to different paths and allow us to easily reschedule those portions whenever necessary.

\section{Conclusions}\label{sec:conclusions}

We presented a detailed description of CTCP, a coded multi-path transport protocol that provides reliability, congestion control, and support for delay sensitive applications. The protocol presented was implemented over UDP to illustrate its performance and benefits. Our experiments indicate that CTCP (single or multiple paths) achieves significantly higher throughput and reliability in faulty and lossy networks than TCP, and is able to load balance the traffic over multiple paths.

\bibliography{ctcp_bib}
\bibliographystyle{IEEEtran}

\end{document}